\documentclass[a4paper,11pt]{article}
\usepackage[latin1]{inputenc}
\usepackage[T1]{fontenc}
\usepackage[english]{babel}
\usepackage{amssymb}
\usepackage{amsfonts, amsmath}
\newtheorem{Theo}{Theorem}
\title{\textbf{The Yang Mills system and cyclic covering of abelian varieties}}
\author{\textbf{A. Lesfari}
\\\emph{Department of Mathematics}
\\\emph{Faculty of Sciences}
\\\emph{University of Choua\"{i}b Doukkali}
\\\emph{B.P. 20, El-Jadida, Morocco}.
\\\emph{E. mail address} : Lesfariahmed@yahoo.fr, lesfari@ucd.ac.ma}
\date{}
\begin{document}
\maketitle
\begin{abstract}
In this paper, we consider a dynamical system related to the Yang-Mills system
for a field with gauge group SU(2). We solve this system in terms of genus two
hyperelliptic functions and we show that it is algebraic completely integrable
in the generalized sense.\\
\emph{Key words.} Integrable systems , Riemann surfaces, Abelian
varieties, Surfaces of general type.\\
\emph{MSC 2000}. 70H06, 14H55, 14K20, 14H70.
\end{abstract}

\section{Introduction}

The problem of finding and integrating hamiltonian systems, has
attracted a considerable amount of attention in recent years.
Beside the fact that many integrable hamiltonian systems have been
on the subject of powerful and beautiful theories of mathematics,
another motivation for its study is: the concepts of integrability
have been applied to an increasing number of physical systems,
biological phenomena, population dynamics, chemical rate
equations, to mention only a few. However, it seems still hopeless
to describe or even to recognize with any facility, those
hamiltonian systems which are integrable, though they are quite
exceptional.\\
The resolution of the well known Korteweg-de-Vries (K-dV) equation
has generated an enormous number of new ideas in the area of
hamiltonian completely integrable systems. It has led to
unexpected connections between mechanics, spectral theory, Lie
algebra theory, algebraic geometry and even differential geometry.
All these connections have generated renewed interest in the
questions around complete integrability of finite and infinite
dimensional systems, ordinary and partial differential equations.
However given a hamiltonian system, it remains often hard to fit
it into any of those general frameworks. But luckily, most of the
problems  possess the mutch richer structure of the so called
algebraic complete integrability (concept introduced et
systematized  by Adler and van Moerbeke). A dynamical system is
algebraic completely integrable in the sense of Adler-van Moerbeke
[1] if it can be linearized on a complex algebraic torus
$\mathbb{C}^{n}/lattice$ (=abelian variety). The invariants (often
called first integrals or constants) of the motion are polynomials
and the phase space coordinates (or some algebraic functions of
these) restricted to a complex invariant variety defined by
putting these invariants equals to generic constants, are
meromorphic functions on an abelian variety. Moreover, in the
coordinates of this abelian variety, the flows (run with complex
time) generated by the constants of the motion are straight lines.
Some results concerning geodesic flow on $SO(4)$ [1,8],
Kowalewski's top [11], Hénon-Heiles system [14],...was obtained.
However, besides the fact that many hamiltonian completely
integrable systems posses this structure, another motivation for
its study which sounds more modern is: algebraic completely
integrable systems come up systematically whenever you study the
isospectral deformation of some linear operator containing a
rational indeterminate. Therefore there are hidden symmetries
which have a group theoretical foundation. The concept of
algebraic complete integrability is quite effective in small
dimensions and has the advantage to lead to global results, unlike
the existing criteria for real analytic integrability, which, at
this stage are perturbation results. In fact, the overwhelming
majority of dynamical systems, hamiltonian or not, are
non-integrable and possess regimes of chaotic behavior in phase
space.\\
In the present paper, we discuss an interesting interaction
between complex geometry and dynamical systems. We shall be
concerned with an integrable system which appears as covering of
another algebraic completely integrable system. The invariant
variety is covering of abelian variety and this system is
algebraic completely integrable in the generalized sense.

\section{The nonlinear Yang Mills equations for a field with gauge group $SU(2)$}

We consider the Yang-Mills system for a field with gauge group
$SU(2):$
$$D_{j}F_{jk}=\partial _{j}F_{jk}+[A_{j},F_{jk}] =0,$$
where $F_{jk}, A_{j}\in T_{e}SU(2), 1\leq j,k\leq 4$ and
$$F_{jk}=\partial _{j}A_{k}-\partial _{k}A_{j}+[ A_{j},A_{k}].$$ The
self-dual Yang-Mills (SDYM) equations is a universal system for
which some reductions include all classical tops from Euler to
Kowalewski (0+1-dimensions), K-dV, Nonlinear Schrödinger,
Sine-Gordon, Toda lattice and N-waves equations (1+1-dimensions),
KP and D-S equations (2+1-dimensions), etc... In the case of
homogeneous double-component field,
\begin{eqnarray}
\partial_{j}A_{k}&=&0,\quad(j\neq
1),\nonumber\\
A_{1}&=&A_{2}=0,\nonumber\\
A_{3}&=&n_{1}U_{1}\in su( 2),\nonumber\\
A_{4}&=&n_{2}U_{2}\in su(2),\nonumber
\end{eqnarray}
where $n_{i}$ are $su(2) $-generators, i.e., they satisfy
commutation relations :
\begin{eqnarray}
n_1&=&[n_2,[n_1,n_2]],\nonumber\\
n_2&=&[n_1,[n_2,n_1]].\nonumber
\end{eqnarray}
The system becomes
$$\partial ^{2}U_{1}+U_{1}U_{2}^{2}=0,$$
$$\partial ^{2}U_{2}+U_{2}U_{1}^{2}=0.$$
By setting
\begin{eqnarray}
U_{j}&=&q_{j},\nonumber\\
\frac{\partial U_{j}}{\partial t}&=&p_{j},\quad j=1,2,\nonumber
\end{eqnarray}
Yang-Mills equations are reduced to hamiltonian system with the
hamiltonian $$H=\frac{1}{2}(
p_{1}^{2}+p_{2}^{2}+q_{1}^{2}q_{2}^{2}).$$ The symplectic
transformation
\begin{eqnarray}
p_{1}&\longleftarrow& \frac{\sqrt{2}}{2} (
p_{1}+p_{2}),\nonumber\\
p_{2}&\longleftarrow& \frac{\sqrt{2}}{2} ( p_{1}-p_{2}),\nonumber\\
q_{1}&\longleftarrow &\frac{1}{2}( \root{4}\of{2})( q_{1}+iq_{2}),\nonumber\\
q_{2}&\longleftarrow&\frac{1}{2}( \root{4}\of{2}) (
q_{1}-iq_{2}),\nonumber
\end{eqnarray}
takes this hamiltonian into
\begin{equation}\label{eqn:euler}
H=\frac{1}{2}(p_{1}^{2}+p_{2}^{2})+\frac{1}{4}q_{1}^{4}
+\frac{1}{4}q_{2}^{4}+\frac{1}{2}q_{1}^{2}q_{2}^{2}.
\end{equation}
We start with the hamiltonian
\begin{equation}\label{eqn:euler}
H=\frac{1}{2}( p_{1}^{2}+p_{2}^{2}+a_{1}q_{1}^{2}+a_{2}q_{2}^{2})
+\frac{1}{4}q_{1}^{4}+\frac{1}{4}a_{3}q_{2}^{4}+\frac{1}{2}a_{4}q_{1}^{2}q_{2}^{2}.
\end{equation}
Note that if $a_1=a_2=0$ and $a_3=a_4=1,$ we obtain the
hamiltonian (1). It has been shown [10] that if $
a_{2}=4a_{1}\equiv 4a, a_{3}=16, a_{4}=6,$ i.e.,
\begin{equation}\label{eqn:euler}
H_1\equiv H=\frac{1}{2}(
p_{1}^{2}+p_{2}^{2})+\frac{a}{2}(q_{1}^{2}+4q_{2}^{2})
+\frac{1}{4}q_{1}^{4}+4q_{2}^{4}+3q_{1}^{2}q_{2}^{2},
\end{equation}
the corresponding system, i.e.,
\begin{eqnarray}
\dot q_{1}&=&p_{1},\nonumber\\
\dot q_{2}&=&p_{2},\\
\dot p_{1}&=&-(a+q_{1}^{2}+6q_{2}^{2})q_{1},\nonumber\\
\dot p_{2}&=&-2(2a+3q_{1}^{2}+8q_{2}^{2})q_{2},\nonumber
\end{eqnarray}
is integrable, the second integral is
\begin{equation}\label{eqn:euler}
H_{2}=aq_{1}^{2}q_{2}+q_{1}^{4}q_{2}+2q_{1}^{2}q_{2}^{3}
-q_{2}p_{1}^{2}+q_{1}p_{1}p_{2},
\end{equation}
but no description of solutions is given. We solve the system (4)
in terms of genus two hyperelliptic functions. When one examines
all possible singularities of the system (4), one finds that it
possible for the variable  $q_1 $ to contain square root terms of
the type $t^{1/2}$, which are strictly not allowed by the so
called Painlevé test (i.e. the general solutions should have no
movable singularities other than poles in the complex plane [5]).

Let $z\equiv(q_{1}, q_{2}, p_{1}, p_{2})\in \mathbb{C}^4$, $t\in
\mathbb{C}$ and $\Delta \subset \mathbb{C}^4$ a non-empty Zariski
open set. By the functional independence of the integrals $H_1,
H_2$, the map
$$\varphi :(H_1, H_2) :
\mathbb{C}^4\longrightarrow\mathbb{C}^2,$$ is submersive, i.e.,
$dH_1(z), dH_2(z)$ are linearly independent on $\Delta$. Let
\begin{eqnarray}
\Omega&=&\varphi(\mathbb{C}^4\backslash \Delta),\nonumber\\
&=&\{b\equiv(b_1,b_2)\in \mathbb{C}^2 : \exists z\in
\varphi^{-1}(b) \mbox{with}\nonumber\\ &&dH_1(z), dH_2(z)
\mbox{linearly dependent}\},\nonumber
\end{eqnarray}
be the set of critical values of $\varphi$. We denote by
$\overline{\Omega}$ the Zariski closure of $\Omega$ in
$\mathbb{C}^2$. The set $\{z\in\mathbb{C}^4 :
\varphi(z)\in\mathbb{C}^2\backslash\overline{\Omega}\}$ is a
non-empty Zariski open set in $\mathbb{C}^4$. Hence this set is
everywhere dense in $\mathbb{C}^4$ for the usual topology. Let $A$
be the complex affine variety defined by
\begin{eqnarray}
A&=&\varphi^{-1}(b),\nonumber\\
&=&\bigcap_{k=1}^{2}\{z\in\mathbb{C}^4:H_k(z)=b_k\}.
\end{eqnarray}
For every $b\equiv(b_1,
b_2)\in\mathbb{C}^2\backslash\overline{\Omega}$, the fibre $A$ is
a smooth affine surface.

\section{Laurent series solutions and algebraic curves}

We show that the system (4) admits Laurent solutions in $t^{1/2}$,
depending on three free parameters: $u, v $ and $w$. These pole
solutions restricted to the surface $A$(6) are parameterized by
two smooth curves $\mathcal{C}_{\varepsilon=\pm i}$(8) of genus 4.

Recall that a system $\dot{z}=f(z)$ is weight-homogeneous with a
weight $\nu_k$ going with each variable $z_k$ if
$$f_k(\lambda^{\nu_i}z_1,\ldots,\lambda^{\nu_n}z_n)=\lambda
^{\nu_k+1}f_k'z_1,\ldots,z_n),$$ for all $\lambda \in \mathbb{C}.$
The system (4) is weight-homogeneous with $q_1, q_2$ having weight
1 and $p_1, p_2$ weight 2, so that $H_1$ and $H_2$ have weight 4
and 5 respectively.

\begin{Theo}
The system (4) admits Laurent solutions in $t^{1/2}$, depending on
3 free parameters: $u, v $ and $w$. These solutions restricted to
the surface $A$(6) are parameterized by two smooth curves
$\mathcal{C}_{\varepsilon=\pm i}$(8) of genus 4.
\end{Theo}
\emph{Proof}. The system (4) possesses 3-dimensional family of
Laurent solutions (principal balances) depending on three free
parameters $u, v $ and $w$. There are precisely two such families,
labeled by $\varepsilon =\pm i,$ and they are explicitly given as
follows
\begin{eqnarray}
q_{1}&=&\frac{1}{\sqrt{t}}( u -\frac{1}{2}u ^{3}t+v t^{2}+u ^{2} (
-\frac{11}{16}u ^{5}+\frac{1}{3}a
u +v )t^{3}\nonumber\\
&&+\frac{u }{4}( \frac{41}{32}u ^{8}-au ^{4}
 +\frac{3}{2}u ^{3}v +\frac{1}{6}a ^{2}-\frac{3\varepsilon
 \sqrt{2}}{2}w
 ) t^{4}+\cdots ),\\
q_{2}&=&\frac{\varepsilon \sqrt{2}}{4t}(1+ u ^{2}t+\frac{1}{3}( 2a
-3u ^{4}) t^{2}+\frac{1}{8}u (
24v -u^{5}) t^{3}-2\varepsilon \sqrt{2}w t^{4}+\cdots ) , \nonumber\\
p_{1}&=&\frac{1}{t\sqrt{t}}( -\frac{1}{2}u -\frac{1}{4}u ^{3}t
+\frac{3}{2}v t^{2}+\frac{5}{2}u ^{2}(
-\frac{11}{16}u ^{5}+\frac{1}{3}a u +v) t^{3}\nonumber\\
&&+\frac{7u }{8}( \frac{41}{32}u ^{8}-au ^{4} +\frac{3}{2}u ^{3}v
+\frac{1}{6}a ^{2}-\frac{3\varepsilon \sqrt{2}}{2}w ) t^{4}
+\cdots ), \nonumber\\
p_{2}&=&\frac{\varepsilon \sqrt{2}}{4t^{2}}( -1+\frac{1}{3}(2a
-3u^{4})t^{2}+\frac{1}{4}u (24v -u ^{5})t^{3}-6\varepsilon
\sqrt{2}w t^{4}+\cdots ). \nonumber
\end{eqnarray}
These formal series solutions are convergent as a consequence of
the majorant method. By substituting these series in the constants
of the motion $H_{1}=b_{1}$ and $H_{2}=b_{2},$ one eliminates the
parameter $w $ linearly, leading to an equation connecting the two
remaining parameters $u$ and $v$ :
\begin{equation}\label{eqn:euler}
2v ^{2}+\frac{1}{6}( 15u ^{4}-8a ) u v -\frac{39}{32}u
^{10}+\frac{7}{6}a u ^{6} +\frac{2}{9}( a ^{2}+9b_{1}) u
^{2}-\varepsilon \sqrt{2}b_{2}=0.
\end{equation}
This defines two smooth curves $\mathcal{C}_\varepsilon $
$(\varepsilon=\pm i)$. Let $g(\mathcal{C}_\varepsilon)$=genus of $
\mathcal{C}_\varepsilon$, $n=\mbox{number of sheets}$ and
$v=\mbox{number of branch points}$. Then by the Riemann-Hurwitz's
formula [17],
$$g(\mathcal{C}_\varepsilon)=-n+1+\frac{v}{2}=-2+1+\frac{10}{2}=4,$$
which finishes the proof of the theorem.

\section{Linearizing the flow in terms of genus two hyperelliptic functions}
\begin{Theo}
The system of differential equations (4) can be integrated in
terms of genus 2 hyperelliptic functions.
\end{Theo}
\emph{Proof}. We set
\begin{eqnarray}
q_2&=&s_1+s_2,\nonumber\\
q_{1}^{2}&=&-4s_1s_2,\nonumber\\
p_2&=&\dot s_1+\dot s_2,\nonumber\\
q_1p_1&=&-2(\dot s_1s_2+s_1\dot s_2).\nonumber
\end{eqnarray}
The latter equation together with the second implies that
$$p_{1}^{2}=-\frac{(\dot s_1s_2+s_1\dot
s_2)^2}{s_1s_2}.$$ In term of these new variables, equations (3)
and (5) take the following form
\begin{eqnarray}
&&\left(s _{1}-s _{2}\right) \left(s _{2}(\dot s_1)^{2}-s
_{1}(\dot s_2)^{2}\right)
\nonumber\\
&&+4s _{1}s _{2}\left( 2s _{1}^{4}+2s _{1}^{3}s _{2}+2s _{1}^{2}s
_{2}^{2}+2s_{1}s _{2}^{3}+2s _{2}^{4}+as _{1}^{2}+as
_{1}s _{2}+as _{2}^{2}\right)\nonumber\\
&&-2b_{1}s _{1}s _{2}=0,\nonumber\\
&&\left(s _{1}-s _{2}\right) \left(s _{2}^{2}(\dot
s_1)^{2}-\allowbreak s
_{1}^{2}(\dot s_2)^{2}\right) \nonumber\\
&&+4s _{1}^{2}s _{2}^{2}\left(s _{1}+s _{2}\right) \left(
a+\allowbreak 2s _{1}^{2}+2s _{2}^{2}\right) +b_{2}s _{1}s _{2}=0
.\nonumber
\end{eqnarray}
These equations are solved linearly for $(\dot s_1)^2$ and $(\dot
s_2)^2$ as
\begin{eqnarray}
(\dot s_1)^2&=&\frac{s_{1}(-8s _{1}^{5}-4a s _{1}^{3}+2b_{1}s
_{1}+b_{2})}{(s_{1}-s _{2})
^{2}},\nonumber\\
(\dot s_2)^2&=&\frac{s _{2}(-8s _{2}^{5}-4as _{2}^{3} +2b_{1}s
_{2}+b_{2})}{(s _{1}-s _{2})^{2}},\nonumber
\end{eqnarray}
which leads immediately to the following equations for $s_{1}$ and
$s_{2}:$
$$\dot{s}_{1}=\frac{ds_{1}}{dt}=\frac{\sqrt{P_{6}(s_{1})}}{s_{1}-s_{2}},$$
$$\dot{s}_{2}=\frac{ds_{2}}{dt}=\frac{\sqrt{P_{6}(s_{2})}}{s_{2}-s_{1}},$$
where $P_6(s)$ is a polynomial of degree 6 of the form
$$
P_{6}(s)=s (-8s ^{5}-4a s ^{3}+2b_{1}s +b_{2}).
$$
These equations can be integrated by the abelian mapping
$$
\mathcal{H}\longrightarrow
Jac(\mathcal{H})=\mathbb{C}^{2}/\Lambda,\quad (p_{1},p_{2})
\longmapsto (\xi _{1},\xi _{2}),
$$
where the hyperelliptic curve $\mathcal{H}$ of genus 2 is given by
the equation $$\zeta^2=P_6(s),$$ $\Lambda $ is the lattice
generated by the vectors $n_{1}+\Omega n_{2},(n_{1},n_{2}) \in
\mathbb{Z}^{2},\Omega $\ is the matrix of period of the curve
$\mathcal{H}$, $p_{1}=(s_{1},\sqrt{P_{6}( s_{1})}),$
$p_{2}=(s_{2},\sqrt{P_{6}(s_{2})})$,
$$\xi _{1}=\int_{p_{0}}^{p_{1}}\omega
_{1}+\int_{p_{0}}^{p_{2}}\omega _{1},$$
$$\xi _{2}=\int_{p_{0}}^{p_{1}}\omega _{2}+\int_{p_{0}}^{p_{2}}\omega
_{2},$$ where $p_{0}$ is a fixed point and $(\omega _{1},\omega
_{2})$\ is a canonical basis of holomorphic differentials on
$\mathcal{H}$, i.e.,
\begin{eqnarray}
\omega_{1}&=&\frac{ds}{\sqrt{P_{6}(s)}}, \nonumber\\
\omega_{2}&=&\frac{sds}{\sqrt{P_{6}(s)}}. \nonumber
\end{eqnarray}
We have
\begin{eqnarray}
\frac{d s _{1}}{\sqrt{P_{6}(s _{1})}}-\frac{ds
_{2}}{\sqrt{P_{6}(s _{2})}}&=&0, \nonumber\\
\frac{s_{1}ds _{1}}{\sqrt{P_{6}(s _{1})}}-\frac{s _{2}ds
_{2}}{\sqrt{P_{6}(s _{2}) }}&=&dt, \nonumber
\end{eqnarray}
and hence the problem can be integrated in terms of genus $2$
hyperelliptic functions of time. This ends the proof of the
theorem.

\section{A five-dimensional system}
We have seen that it possible for the variables $q_1$ and $p_1$ to
contain square root terms of the type $\sqrt{t}$, which are
strictly not allowed by the Painlevé test. However, these terms
are trivially removed by introducing some new variables $
z_1,\ldots,z_5$, which restores the Painlevé property to the
system. Indeed, let
\begin{equation}\label{eqn:euler}
\varphi:A\longrightarrow\mathbb{C}^5,\quad(q_1,q_2,p_1,p_2)\longmapsto(z_1,z_2,z_3,z_4,z_5),
\end{equation}
be a morphism on the affine variety $A(6)$ where $z_1,\ldots,z_5$
are defined as
$$z_1=q_{1}^{2},\quad z_2=q_{2},\quad z_3=p_{2},\quad
z_4=q_{1}p_{1},\quad z_5=2q_{1}^{2}q_{2}^{2}+p_{1}^{2}.$$ The
morphism (9) maps the vector field (4) into the system
\begin{eqnarray}
\dot{z}_1&=&2z_4,\nonumber\\
\dot{z}_2&=&z_3,\nonumber\\
\dot{z}_3&=&-4a z_{2}-6z_{1}z_{2}-16z_{2}^{3} ,\\
\dot{z}_4&=&-az_{1}-z_{1}^{2}-8z_{1}z_{2}^{2}+z_{5},\nonumber\\
\dot{z}_5&=&-8z_{2}^{2}z_{4}-2az_{4}-2z_{1}z_{4}+4z_{1}z_{2}z_{3},\nonumber
\end{eqnarray}
in five unknowns having three quartic invariants
\begin{eqnarray}
F_1&=&\frac{1}{2}z_{5}+2z_{1}z_{2}^{2}+\frac{1}{2}z_{3}^{2}
+\frac{1}{2}az_{1}+2az_{2}^{2}+\frac{1}{4}z_{1}^{2}+4z_{2}^{4},\nonumber\\
F_2&=&az_{1}z_{2}+z_{1}^{2}z_{2}+4z_{1}z_{2}^{3}-z_{2}z_{5}+z_{3}z_{4} ,\\
F_3&=&z_{1}z_{5}-2z_{1}^{2}z_{2}^{2}-z_{4}^{2}.\nonumber
\end{eqnarray}
This system is completely integrable and the hamiltonian structure
is defined by the Poisson bracket
$$\left\{ F,H\right\} =\left\langle \frac{\partial F}{\partial z},
J\frac{\partial H}{\partial z}\right\rangle
=\sum_{k,l=1}^{5}J_{kl}\frac{\partial F}{\partial
z_{k}}\frac{\partial H}{\partial z_{l}},$$ where
$$\frac{\partial H}{\partial
z}=(\frac{\partial H}{\partial z_{1}},\frac{\partial H}{\partial
z_{2}},\frac{\partial H}{\partial z_{3}},\frac{\partial
H}{\partial z_{4}},\frac{\partial H}{\partial z_{5}})^\top,$$ and
$$J=\left[\begin{array}{ccccc}
0&0&0&2z_1&4z_4\\
0&0&1&0&0\\
0&-1&0&0&-4z_1z_2\\
-2z_1&0&0&0&2z_5-8z_1z_{2}^{2}\\
-4z_4&0&4z_1z_2&-2z_5+8z_1z_{2}^{2}&0
\end{array}\right],$$
is a skew-symmetric matrix for which the corresponding Poisson
bracket satisfies the Jacobi identities. The system (10) can be
written as
$$\dot{z}=J\frac{\partial
H}{\partial z},\quad z=(z_{1},z_{2},z_{3},z_{4},z_{5})^\top,$$
where $H=F_{1}.$ The second flow commuting with the first is
regulated by the equations
$$\dot{z}=J\frac{\partial F_{2}}{\partial z},\quad
z=(z_{1},z_{2},z_{3},z_{4},z_{5})^\top,$$ and is written
explicitly as
\begin{eqnarray}
\dot{z}_1&=&2z_{1}z_{3}-4z_{2}z_{4},\nonumber\\
\dot{z}_2&=&z_4,\nonumber\\
\dot{z}_3&=&z_{5}-8z_{1}z_{2}^{2}-az_{1}-z_{1}^{2} ,\nonumber\\
\dot{z}_4&=&-2az_{1}z_{2}-4z_{1}^{2}z_{2}-2z_{2}z_{5},\nonumber\\
\dot{z}_5&=&-4az_{2}z_{4}-4z_{1}z_{2}z_{4}-16z_{2}^{3}z_{4}-2z_{3}z_{5}
+8z_{1}z_{2}^{2}z_{3}.\nonumber
\end{eqnarray}
These vector fields are in involution, i.e.,
$$\{F_1,F_2\}=\langle \frac{\partial F_{1}}{\partial z},J\frac{\partial F_{2}}{\partial
z}\rangle=0,$$ and the remaining one is Casimir, i.e.,
$$J\frac{\partial F_{3}}{\partial z}=0.$$
Let B be the complex affine variety defined by
\begin{equation}\label{eqn:euler}
B=\bigcap_{k=1}^{2}\{z:F_k(z)=c_k\}\subset\mathbb{C}^5,
\end{equation}
for generic $( c_{1},c_{2},c_{3}) \in \mathbb{C}^{3}$. We have shown in [18], that\\
\textbf{a)} The the system (10) can be integrated in genus 2
hyperelliptic
functions.\\
\textbf{b)} The system (10) possesses Laurent series solutions
which depend on 4 free parameters : $\alpha, \beta, \gamma $ and
$\theta:$
\begin{eqnarray}
z_{1}&=&\frac{1}{t}(\alpha -\alpha ^{2}t+\beta t^{2}
+\frac{1}{6}\alpha (3\beta -9\alpha ^{3}+4a\alpha)t^{3}+\gamma t^{4}+\cdots ),\nonumber\\
z_{2}&=&\frac{\varepsilon \sqrt{2}}{4t}(1+\alpha
t+\frac{1}{3}(-3\alpha ^{2}+2a)t^{2}+\frac{1}{2}(3\beta -\alpha
^{3})t^{3}-2\varepsilon \sqrt{2}\theta t^{4}+\cdots), \nonumber\\
z_{3}&=&\frac{\varepsilon
\sqrt{2}}{4t^{2}}(-1+\frac{1}{3}(-3\alpha ^{2}
+2a)t^{2}+(3\beta -\alpha ^{3})t^{3}-6\varepsilon \sqrt{2}\theta t^{4}+\cdots),\\
z_{4}&=&\frac{1}{2t^{2}}(-\alpha +\beta t^{2}+\frac{1}{3}\alpha
(3\beta -9\alpha ^{3}
+4a\alpha)t^{3}+3\gamma t^{4}+\cdots), \nonumber\\
z_{5}&=&\frac{1}{t}(-\frac{1}{3}a \alpha +\alpha ^{3}-\beta
+( 3\alpha ^{4}-a\alpha ^{2} -3\alpha \beta)t \nonumber\\
&&+( 4\varepsilon \sqrt{2}\alpha \theta +2\gamma
+\frac{8}{3}a\alpha ^{3} -\frac{1}{3}a \beta -\alpha ^{2}\beta
-3\alpha ^{5}-\frac{4}{9}a^2\alpha)t^{2}+\cdots), \nonumber
\end{eqnarray}
with $\varepsilon=\pm i.$ These meromorphic solutions restricted
to the surface B(12) are parameterized by two isomorphic smooth
hyperelliptic curves $\mathcal{H}_{\varepsilon=\pm i}$ of genus 2
:
\begin{equation}\label{eqn:euler}
\beta ^{2}+\frac{2}{3}(3\alpha ^{2}-2a)\alpha \beta -3\alpha ^{6}+
\frac{8}{3}a\alpha ^{4}+\frac{4}{9}(a^{2}+9c_{1})\alpha
^{2}-2\varepsilon \sqrt{2}c_{2}\alpha +c_{3}=0,
\end{equation}
\textbf{c)} The variety $B$(12) is embedded in $\mathbb{P}^{15}$
and generically is the affine part of an abelian surface
$\widetilde{B},$ more precisely the jacobian of a genus 2 curve.
The reduced divisor at infinity
$$\widetilde{B}\backslash B=\mathcal{H}_i+\mathcal{H}_{-i},$$
consists of two smooth isomorphic genus 2 curves
$\mathcal{H}_\varepsilon $(14), that intersect in only one point
at which they are tangent to each other. The system of
differential equations (10) is algebraically completely integrable
and
the corresponding flows evolve on $\widetilde{B}.$\\

Observe that the reflection $\sigma$ on the affine variety B
amounts to the flip $$\sigma :(z_1,z_2,z_3,z_4,z_5)\longmapsto
(z_1,z_2,-z_3,-z_4,z_5),$$ changing the direction of the commuting
vector fields. It can be extended to the (-Id)-involution about
the origin of $\mathbb{C}^2$ to the time flip $(t_1,t_2)\mapsto
(-t_1,-t_2)$ on $\widetilde{B}$, where $t_{1}$ and $t_{2}$ are the
time coordinates of each of the flows $X_{{F}_{1}}$ and
$X_{{F}_{2}}.$ The involution $\sigma $ acts on the parameters of
the Laurent solution (13) as follows
$$\sigma :(t,\alpha,\beta,\gamma,\theta,\varepsilon)\longmapsto
(-t,-\alpha,-\beta,-\gamma,-\theta,-\varepsilon),$$ interchanges
the curves $\mathcal{H}_{\varepsilon =\pm i}$(14). Geometrically,
this involution interchanges $\mathcal{H}_{i}$ and
$\mathcal{H}_{-i},$ i.e., $\mathcal{H}_{-i}=\sigma
\mathcal{H}_{i}.$\\

The asymptotic solution (7) can be read off from (13) and the
change of variable : $$q_1=\sqrt{z_1},\quad q_2=z_2,\quad
p_1=z_4/q_1,\quad p_2=z_3.$$ The function $z_1$ has a simple pole
along the divisor $\mathcal{H}_i+\mathcal{H}_{-i}$ and a double
zero along a hyperelliptic curve of genus 2 defining a double
cover of $\widetilde{B}$ ramified along
$\mathcal{H}_i+\mathcal{H}_{-i}.$

\section{Generalized algebraic completely integrable system}

Applying the method explained in Piovan [22], we show that the
invariant variety A(6) can be completed as a cyclic double cover
$\overline{A}$ of the jacobian of a genus curve, ramified along a
divisor $\mathcal{H}_i+\mathcal{H}_{-i}$ where $\mathcal{H}_i$ and
$\mathcal{H}_{-i}$ are two isomorphic hyperelliptic curves (14) of
genus 2 that intersect in only one point at which they are tangent
to each other. Moreover, $\overline{A}$ is smooth except at the
point lying over the singularity (of type $A_3$) of
$\mathcal{H}_i+\mathcal{H}_{-i}$ and the resolution
$\widetilde{A}$ of $\overline{A}$ is a surface of general type
with invariants : Euler characteristic of $\widetilde{A}=$
$\mathcal{X}(\widetilde{A})=1$ and geometric genus of
$\widetilde{A}=$ $p_g(\widetilde{A})=2.$ Consequently, the system
(4) is algebraic completely integrable in the generalized sense.

\begin{Theo}
The invariant surface $A$(6) can be completed as a cyclic double
cover $\overline{A}$ of the abelian surface $\widetilde{B}$ (the
jacobian of a genus 2 curve), ramified along the divisor
$\mathcal{H}_{i}+\mathcal{H}_{-i}.$ The system (4) is algebraic
complete integrable in the generalized sense. Moreover,
$\overline{A}$ is smooth except at the point lying over the
singularity (of type $A_3$) of $\mathcal{H}_{i}+\mathcal{H}_{-i}$
and the resolution $\widetilde{A}$ of $\overline{A}$ is a surface
of general type with invariants : $\mathcal{X}(\widetilde{A})=1$
and $p_g(\widetilde{A})=2.$
\end{Theo}
\emph{Proof}. We have shown that the morphism $\varphi$ (9) maps
the vector field (4) into an algebraic completely integrable
system (10) in five unknowns and the affine variety $A$ (6) onto
the affine part $B$ (12) of an abelian variety $\widetilde{B}$
(more precisely the jacobian of a genus 2 curve with
$\widetilde{B}\backslash B=\mathcal{H}_{i}+\mathcal{H}_{-i}$).
Observe that $\varphi:A\longrightarrow B$ is an unramified cover.
The curves $\mathcal{C}_\varepsilon $ (8) play an important role
in the construction of a compactification $\overline{A}$ of $A.$
Let us denote by $G$ a cyclic group of two elements $\{-1,1\}$ on
$$V_\varepsilon ^j=U_\varepsilon ^j \times \{\tau \in \mathbb{C} :
0<|\tau|<\delta\},$$ where $\tau=t^{1/2}$ and $ U_\varepsilon ^j$
is an affine chart of $\mathcal{C}_\varepsilon$ for which the
Laurent solutions (7) are defined. The action of $G$ is defined by
$$(-1)\circ (u,v,\tau)=(-u,-v,-\tau),$$ and is without fixed points
in $V_\varepsilon ^j.$ So we can identify the quotient
$V_\varepsilon ^j / G$ with the image of the smooth map
$h_\varepsilon ^j :V_\varepsilon ^j\rightarrow A$ defined by the
expansions (7). We have $$(-1,1).(u,v,\tau)=(-u,-v,\tau),$$ and
$$(1,-1).(u,v,\tau)=(u,v,-\tau),$$ i.e., $G\times G$ acts separately
on each coordinate. Thus, identifying $V_\varepsilon ^j/G^2$ with
the image of $\varphi\circ h_\varepsilon ^j$ in $B.$ Note that
$A_\varepsilon ^j=V_\varepsilon ^j/G$ is smooth (except for a
finite number of points) and the coherence of the $A_\varepsilon
^j$ follows from the coherence of $V_\varepsilon ^j$ and the
action of $G.$ Now by taking $A$ and by gluing on various
varieties $A_\varepsilon ^j\backslash \{\mbox{some points}\},$ we
obtain a smooth complex manifold $\widehat{A}$ which is a double
cover of the abelian variety $\widetilde{B}$ ramified along
$\mathcal{H}_i+\mathcal{H}_{-i},$ and therefore can be completed
to an algebraic cyclic cover of $\widetilde{B}.$ To see what
happens to the missing points, we must investigate the image of
$\mathcal{C}_\varepsilon \times\{0\}$ in $\cup A_\varepsilon ^j.$
The quotient $\mathcal{C}_\varepsilon \times\{0\}/G$ is
birationally equivalent to the smooth hyperelliptic curve
$\Gamma_\varepsilon$ of genus 2 :
$$2w^2+\frac{1}{6}(15z^2-8a)zw+z(-\frac{39}{32}z^5+\frac{7}{6}az^3
+\frac{2}{9}(a^2+9b_1)z-\varepsilon \sqrt{2}b_2)=0,$$ where $w=uv,
z=u^2.$ The curve $\Gamma_\varepsilon$ is birationally equivalent
to $\mathcal{H}_\varepsilon.$ The only points of
$\mathcal{C}_\varepsilon$ fixed under $(u,v)\mapsto (-u,-v)$ are
the two points at $\infty,$ which correspond to the ramification
points of the map
$$
\mathcal{C}_\varepsilon \times\{0\}\overset{2-1}{\longrightarrow
}\Gamma_\varepsilon :(u,v)\longmapsto(z,w),
$$
and coincides with the points at $\infty$ of the curve
$\mathcal{H}_\varepsilon.$ Then the variety $\widehat{A}$
constructed above is birationally equivalent to the
compactification $\overline{A}$ of the generic invariant surface
$A.$ So $\overline{A}$ is a cyclic double cover of the abelian
surface $\widetilde{B}$ (the jacobian of a genus 2 curve) ramified
along the divisor $\mathcal{H}_i+\mathcal{H}_{-i},$ where
$\mathcal{H}_i$ and $\mathcal{H}_{-i}$ intersect each other in a
tacnode. It follows that The system (4) is algebraic complete
integrable in the generalized sense. Moreover, $\overline{A}$ is
smooth except at the point lying over the singularity (of type
$A_3$) of $\mathcal{H}_i+\mathcal{H}_{-i}.$ In term of an
appropriate local holomorphic coordinate system $(x,y,z),$ the
local analytic equation about this singularity is $x^4+y^2+z^2=0.$
Now, let $\widetilde{A}$ be the resolution of singularities of
$\overline{A},$ $\mathcal{X}(\widetilde{A})$ be the Euler
characteristic of $\widetilde{A}$ and $p_g(\widetilde{A})$ the
geometric genus of $\widetilde{A}.$ Then $\widetilde{A}$ is a
surface of general type with invariants :
$\mathcal{X}(\widetilde{A})=1$ and $p_g(\widetilde{A})=2.$ This
concludes the proof of the theorem.

\end{document}